\documentclass[aps,twocolumn,showpacs,float]{revtex4}

\usepackage{amsmath}
\usepackage{graphicx}
\usepackage{float}
\floatstyle{boxed}
\usepackage{bm}

\begin{document}

\bibliographystyle{prsty}
\author{Eugene M. Chudnovsky$^{1}$ and Reem Jaafar$^2$}
\affiliation{$^{1}$Department of Physics and Astronomy, Lehman College and Graduate School, The City University of New York, 250 Bedford Park Boulevard West, Bronx, NY 10468-1589\\
$^{2}$Department of Mathematics, Engineering and Computer Science, LaGuardia Community College, The City University of New York, 31-10 Thomson Avenue, Long Island City, NY 11101}
\date{\today}

\begin{abstract}
We show that the magnetic moment of a composite multiferroic torsional oscillator can be switched by the electric field. The 180$^o$ switching arises from the spin-rotation coupling and is not prohibited by the different symmetry of the magnetic moment and the electric field as in the case of a stationary magnet. Analytical equations describing the system have been derived and investigated numerically. Phase diagrams showing the range of parameters required for the switching have been obtained.  
\end{abstract}
\pacs{85.85.+j; 77.80.Fm; 75.78.Jp; 77.65.-j}

\title{Electromechanical Magnetization Switching}

\maketitle

\section{Introduction}

There has been enormous interest in investigating the switching of the magnetization of a nanomagnet by the spin-polarized electric current or by the electric field as possible routes to increasing computer speed. Spin transfer torque random access memory (STT-RAM) devices have been successfully engineered as cache memory and have entered the market. Large effort has been developed to achieve fast switching of the magnetic moment by the electric field in magnets interfaced with piezoelectric materials and in multiferroic systems \cite{bibbar08natmat,chumarhol08natmat,wanliuren09advphy,hertraash11prl,ferzahost12prl,luxia14prb}. This latter task is difficult to achieve due the linear relation, ${\bf M} \propto {\bf E}$, between the magnetic moment ${\bf M}$ and the electric field ${\bf E}$ is prohibited by symmetry. It prompted researchers to look for, e.g., $90$-degree rotation of the magnetization through the effect of the electric field on the magnetic anisotropy, or consider combined effects of the electric field and the exchange bias in layered systems. 

In this paper we propose a nanoelectromechanical system (NEMS) for switching of the magnetic moment by the electric field that is based upon a different principle. It follows from the observation that symmetry permits linear relation between ${\bf M}$ and the angular velocity ${\bm \Omega}$ of the rotation. The latter can be generated by a pulse of the electric field applied to a multiferroic particle attached to a torsional cantilever, or by the electric field pulse applied to a ferroelectric cantilever containing a magnetic particle. According to the Larmor theorem, in the coordinate frame of a nanomagnet, torsional vibrations are equivalent to the ac magnetic field \cite{chu04prl,chugarsch05prb}. The latter is known to be capable of switching the magnetic moment, see, e.g., Ref.  \onlinecite{CGC-PRB2013} and references therein. The advantage of mechanical vibrations over the ac magnetic field is that they can be localized at the nanoscale. This permits miniaturization down to molecular dimensions as has been demonstrated by recent measurements of a single molecular spin in a NEMS obtained by grafting of a single-molecule magnet on a carbon nanotube \cite{ganklyrub13NatNano,ganklyrub13acsNano}. 

\begin{figure}
\includegraphics[width=90mm]{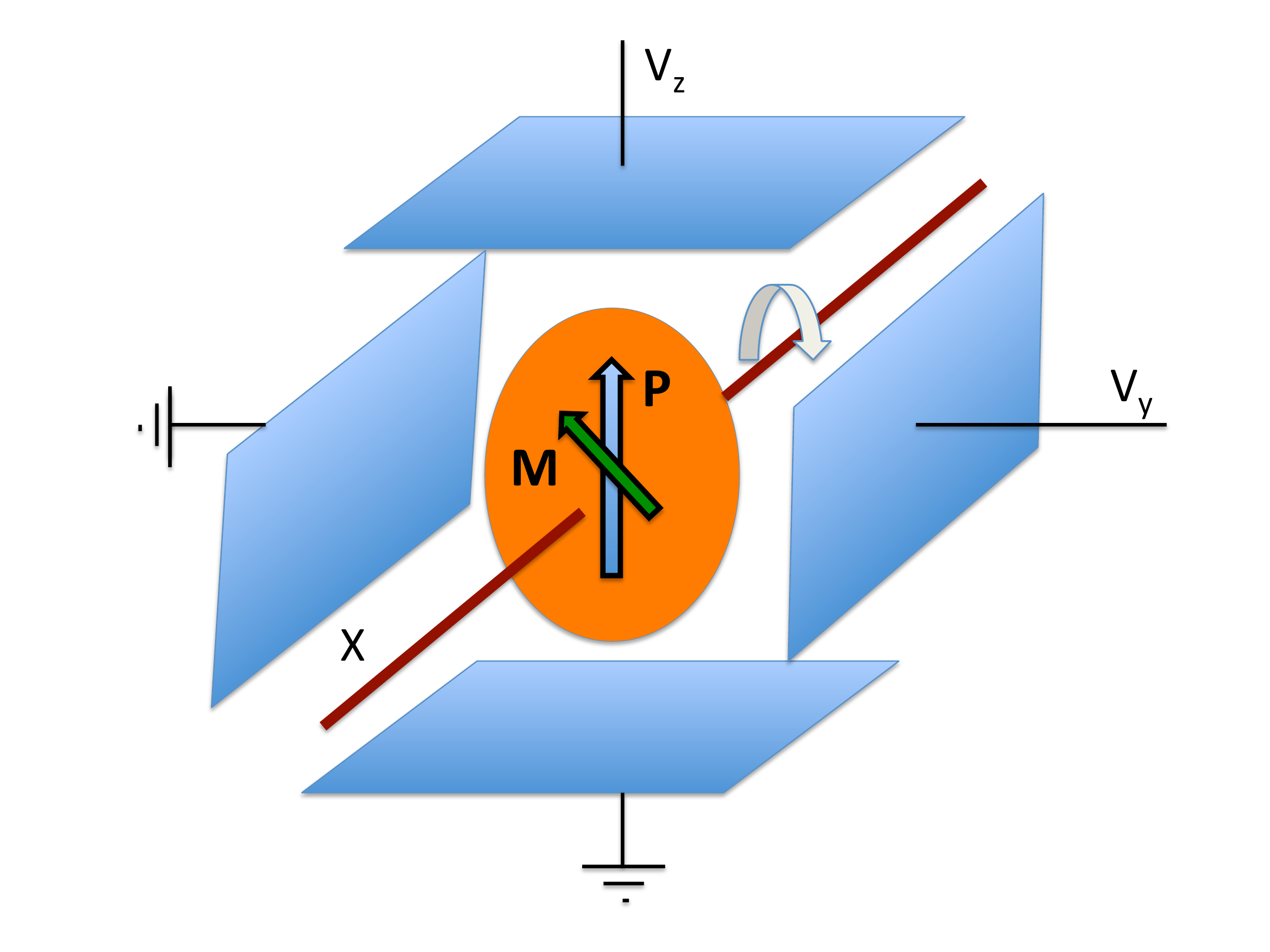}
\caption{Schematic representation of the geometry of the model studied in the paper. A particle with a magnetic moment ${\bf M}$ and electric polarization ${\bf P}$  is a part of a torsional oscillator rotating about the $X$ axis. The system is placed inside a pair of capacitors. The reversal of the magnetic moment is generated by pulse voltages, $V_x$ and $V_y$, on the capacitors.}
\label{geometry}
\end{figure}
Coupling of mechanical vibrations to the magnetization of a thin film deposited on a small cantilever has been studied before \cite{Kovalev-APL2003}. In a multidomain magnetic film it is related to the motion of domain walls \cite{Wallis-APL2006,jaachugar09prb,limimtwal14epl}.  In a single-domain system the magnetomechanical magnetization reversal was shown to have strong sensitivity to the parameters \cite{Kovalev-PRL2005}. The concept of spin-rotation coupling  sheds new light on the origin of the observed stochasticity.  Circularly polarized ac field switches the magnetic moment in a predictable manner via the consecutive absorption of photons of one helicity (i.e., one spin projection) while a linearly polarized field consists of photons of different helicity that are absorbed intermittently, raising or decreasing the corresponding projection of the magnetic moment. Although torsional vibrations cannot be reduced to either linear or circular polarized ac field, understanding of how the ac field reverses the magnetic moment \cite{CGC-PRB2013} allows one to design the optimal geometry in which rotations induced by the electric field can do the job.  

The equivalence of mechanical rotations to the ac magnetic field in the rotating frame makes the problem of the electromechanical magnetization reversal uniquely defined and free of any unknown coupling constants. Interaction of the magnetic moment with torsional oscillations is completely determined by the Larmor theorem and by the dynamics of the total angular momentum \cite{chu14springer}, spin plus mechanical, in the presence of the mechanical torque generated by the electric field. The paper is organized as follows. The system under consideration and the model are described in Section \ref{model} where dynamical equations for the magnetic moment and mechanical rotations caused by the time-dependent electric field are derived. Numerical solution of the equations and switching phase diagrams are presented in Section \ref{numerical}. Estimates for real systems and possible applications of our results are discussed in Section \ref{discussion}.

\section{The model and dynamics of the system}\label{model}
\subsection{Model}
We consider the system schematically represented in Fig. \ref{geometry}. Magnetic energy of the uniaxial ferromagnetic particle in its own coordinate frame is given by
\begin{equation}
E_M = -KVm_z^2 - V{\bf m} \cdot {\bf H},
\end{equation}
where $K$ is the magnetic anisotropy constant, $V$ is the volume of the particle, ${\bf H}$ is the magnetic field in the frame of the particle, and ${\bf m} = {\bf M}/V$ is its constant-length magnetization. The body-frame $Z$-axis in this formula coincides with the easy magnetization direction. The particle is allowed to rotate mechanically around the $X$-axis, with $\rho$ being the angle that the easy axis makes in the $YZ$-plane with the stationary $Z$ direction in the laboratory frame. 

We shall assume that, in addition to the magnetic moment, the system also possesses the electric polarization characterized by the vector ${\bf P}$ directed along the anisotropy axis of the particle. This can be achieved by either using a composite multiferroic material or by interfacing the magnet with the ferroelectric. Interaction of the electric polarization with the crystal lattice, being of Coulomb origin, is strong, providing that the orientation of the particle in space follows the orientation of ${\bf P}$. On the contrary, interaction of ${\bf M}$ with the mechanical rotation of the particle is of the relativistic spin-orbit origin. Consequently, the magnetization vector  does not necessarily follow the rotations of the anisotropy axis if the latter are sufficiently fast. 

Mechanical rotation of the particle in our model is achieved due to the interaction of ${\bf P}$ with the electric field ${\bf E}$,
\begin{equation}
E_{P} = - {\bf P} \cdot {\bf E}.
\end{equation}
No interaction of ${\bf M}$ with ${\bf P}$ in addition to the magnetic anisotropy is assumed. The effective coupling between ${\bf M}$ and ${\bf E}$ is provided by the rotation of the magnetic anisotropy axis that follows the direction of ${\bf P}$. The easiest way to take this effect into account is by noticing that according to the Larmor theorem the rotation with the angular velocity ${\bm \Omega} = (d\rho/dt)\hat{\bf x}$ generates the effective magnetic field 
\begin{equation}
{\bf H}_R  = \frac{1}{\gamma}\frac{d\rho}{dt}\hat{\bf x}
\end{equation}
in the coordinate frame of the rotating particle. In what follows we shall assume that no external magnetic field is applied. Correspondingly, ${\bf H}_R$ is the only field felt by the particle in its own coordinate frame. 

While the easy magnetization axis in our model follows the direction of the electric polarization ${\bf P}$, the latter does not necessarily follow the direction of the electric field ${\bf E}$ because of the finite moment of inertia of the rotator.  We shall now proceed to the derivation of coupled equations for the magnetic moment and mechanical rotation caused by the application of the time-dependent electric field.

\subsection{Dynamical equation for the magnetic moment in the rotating frame}
The equation of motion for ${\bf s}={\bf m}/|{\bf m}|$ in the coordinate frame rigidly coupled to the particle is \cite{Lectures}
\begin{equation}\label{LL}
\frac{d\mathbf{{s}}}{dt}  = \gamma\left[\mathbf{s}\times {\bf H}_{\rm eff}\right] 
 - \alpha\gamma\left[\mathbf{s}\times\left[\mathbf{s}\times{\bf H}_{\rm eff}\right]\right],
\end{equation}
where  $\gamma$ is the gyromagnetic ratio, $\alpha$ is the dimensionless damping coefficient, and $ {\bf H}_{\rm eff}$ is the effective magnetic field given by
\begin{equation}
{\bf H}_{\rm eff} = -\frac{1}{V}\frac{\partial E_M}{\partial {\bf m}} = {\bf H}_A + {\bf H}_R,
\end{equation}
with ${\bf H}_A = 2Km_z\hat{\bf z}$ being the effective field due to the magnetic anisotropy.

Introducing the frequency of the ferromagnetic resonance, $\omega_{\rm FMR} = 2\gamma Km$, and the dimensionless time $t' = \omega_{\rm FMR}t$, Eq. (\ref{LL}) can be written as
\begin{eqnarray}\label{eq:LLE-Rotating_frame}
\frac{d\mathbf{{s}}}{dt'} & = &\left[\mathbf{s}\times\left(s_z\hat{\bf z}+\frac{d{\rho}(t')}{dt'}\hat{\bf x}\right)\right] \nonumber \\
& -& \alpha\left[\mathbf{s}\times\left[\mathbf{s}\times\left(s_z\hat{\bf z}+\frac{d{\rho}(t')}{dt'}\hat{\bf x}\right)\right]\right].
\end{eqnarray}
At small damping, $\alpha \ll 1$, the last term in Eq. (\ref{eq:LLE-Rotating_frame}) can be replaced with $-\alpha[{\bf s} \times d{\bf s}/dt']$. 

Writing ${\bf s}$ in spherical coordinates in the body frame of the rotator, ${\bf s} = (sin\theta\cos\phi, \sin\theta\sin\phi,\cos\theta)$, one obtains the following equations for the angles describing the direction of the magnetization with respect to the fixed axes inside the rotating particle
\begin{align}
\frac{d\theta}{dt'} = \frac{d\rho}{dt'}\sin\phi -\alpha \sin\theta \cos\theta +\alpha \frac{d\rho}{dt'} \cos\theta \cos\phi
\label{theta-eq-rot-alpha}.
\end{align}
\begin{align}
\frac{d\phi}{dt'} = -\cos\theta + \frac{d\rho}{dt'} \cos\phi \cot\theta - \alpha \frac{d\rho}{dt'} \frac{\sin\phi}{\sin\theta} 
\label{phi-eq-rot-alpha}.
\end{align}

Before studying rotations generated by the electric field, one can ask what $\rho(t)$ would provide the optimum magnetization reversal. In the case of the ac field such optimum reversal occured in one plane, $\phi = {\rm const}$ \cite{CGC-PRB2013}. Setting $\phi = \varphi_0$ in Eqs. (\ref{theta-eq-rot-alpha}) and (\ref{phi-eq-rot-alpha}) one obtains at $\alpha = 0$ 
\begin{equation}\label{const phi}
 \frac{d{\rho}}{dt'} = \frac{1}{\sin \varphi_0}\frac{d{\theta}}{dt'} = \frac{\sin \theta}{\cos \varphi_0}.
\end{equation}
Integration gives
\begin{equation}
\tan \frac{\theta}{2} = e^{t' \tan \varphi_0}
\end{equation}
which corresponds to the switching of the magnetization from $\theta = 0$ at $t' = -\infty$ to $\theta = \pi$ at $t' = +\infty$. According to Eq.\ (\ref{const phi}) the function $\rho(t')$ that generates such  behavior is given by
\begin{equation}
\rho =\frac{2\arctan [\exp (t' \tan \varphi_0)]}{\sin \varphi_0}.
\end{equation}
It describes the rotation that exponentially turns on at $\rho = 0$ and dies out at $\rho = \pi/ \sin \varphi_0$, thus, effectively a characteristic time interval $\Delta t \sim 1/(\omega_{FMR}\tan \varphi_0)$. This kind of switching requires the rotation of the magnet by a significant angle. Curiously, if $\varphi_0 = 30^o$, the $360^o$ rotation does the job. It provides a good hint into the speed and amplitude of the rotations that should be implemented with the help of the electric field pulse to achieve magnetization reversal. 

\subsection{Dynamical equation for the torsional oscillator}
Modeling of the effect of the field requires knowledge of the equation of motion of the torsional oscillator that must be solved together with the dynamical equations for the magnetization, Eqs. (\ref{theta-eq-rot-alpha}) and (\ref{phi-eq-rot-alpha}). For rotations about the $X$-axis it follows from the equation 
\begin{equation}\label{torsional}
\frac{dJ_{x}}{dt}={\tau}_{x},
\end{equation}
 where ${\bf J}$ is the total angular momentum and ${\bm \tau}$ is the mechanical torque. The total angular momentum consists of the spin angular momentum, $\hbar{\bf S} = {\bf M}/\gamma$ and the mechanical angular momentum, $I_r{d\rho}/{dt}$,
\begin{equation}\label{momentum}
{J}_{x}=\hbar S_x+I_r\frac{d\rho}{dt},
\end{equation}
where $I_r$ is the moment of inertia of the rotator. 

When the rotation is generated by the electric field ${\bf E}$ acting on the electric polarization ${\bf P}$, the potential energy of the rotator consists of the elastic part and the part generated by the interaction with the field,
\begin{equation}\label{U}
 U = E_E + E_P = \frac{1}{2}I_r\omega_r^2 \rho^2- {\bf P} \cdot {\bf  E}.
\end{equation}
Here $\omega_r$ is the resonance frequency of the torsional oscillator. Note the proportionality of the elastic energy to $\rho^2$. In nanowires it is justified even for large rotation angles \cite{Weinberger-2010}. The torque is given by 
\begin{equation}\label{torque}
{\tau}_{x}=-\frac{\partial U}{\partial \rho}=-I_r\omega_r^2\rho + \frac{d({\bf P}\cdot{\bf E})}{d\rho},
\end{equation}
where ${\bf P} = P(0,-\sin\rho, \cos\rho)$. 

For, e.g., the electric field directed along the $Y$-axis, ${\bf E} = (0,E,0)$, one obtains from Eq.\ (\ref{torsional}) the following dimensionless torque equation 
\begin{equation}\label{rho-eq-1}
\frac{d^2\rho}{dt'^2} +\omega_r'\eta \frac{d\rho}{dt'}+\omega_r'^2[\rho + p(t')\cos\rho]= -\frac{1}{I_r'}\frac{d}{dt'}(\sin\theta \cos \phi),
\end{equation}
where 
\begin{equation}
\omega_r'=\frac{\omega_r}{\omega_{FMR}}, \quad I_r'=\frac{I_r\omega_{FMR}}{\hbar S}, \quad p(t') = \frac{PE(t')}{I_r\omega_r^2}.
\end{equation}
This case corresponds to the utilization of only the $Y$-capacitor shown in Fig.\ \ref{geometry}. Notice that we have introduced the mechanical damping term in Eq.\ (\ref{rho-eq-1}), proportional to the damping parameter $\eta$. 

For torsional oscillations about a fixed axis, induced by a pulse of the linearly polarized electric field, as in Eq.\ (\ref{rho-eq-1}), the problem resembles the problem with a linearly polarized ac magnetic field for which the chaotic behavior of ${\bf s}$ should be expected. It makes sense, therefore, to consider a two component electric field, ${\bf E} = (0,E_y,E_z)$, generated by two pairs of capacitors shown in Fig. \ref{geometry}. One has to find the optimal ${\bf E}(t)$ that reverses the magnetization with the initial conditions $E = 0$, $\theta = \rho = 0$, at $t =0$. The most promising case turns out to be the case of a rotating field: ${\bf E} = E(t)(0,-\sin \omega' t', \cos \omega' t')$ of varying amplitude $E(t)$. Then ${\bf P} \cdot {\bf E} = PE(t)\cos(\rho - \omega' t')$ and the torque on the magnetic particle is
\begin{equation}
\tau_x  = -I_r\omega_r^2\rho -PE(t)\sin(\rho -\omega' t')
\end{equation}
Instead of Eq.\ (\ref{rho-eq-1}) one obtains
\begin{eqnarray}\label{eq-rho-2}
&&\frac{d^2\rho}{dt'^2} +\omega_r'\eta \frac{d\rho}{dt'}+\omega_r'^2\left[\rho + p(t')\sin(\rho-\omega' t')\right] \nonumber  \\
&& = -\frac{1}{I_r'}\frac{d}{dt'}(\sin\theta \cos \phi)
\end{eqnarray}
The right-hand-side of this equation represents the Einstein - de Haas effect, that is, the back effect of the change in the magnetization on the mechanical motion of the oscillator. It is small when parameter $I'_r$, representing the ratio of the mechanical angular momentum and the spin angular momentum, is large. Notice that a stationary electric field directed along the stationary $Z$-axis in the laboratory frame corresponds to $\omega' = 0$. In this case $\rho = 0$, ${\bf s} = {\rm const}$ is the solution of the equations in accordance with the expectation that such a field does not generate mechanical rotation. 

\section{Numerical results} \label{numerical}

\begin{figure}
\includegraphics[width=80mm]{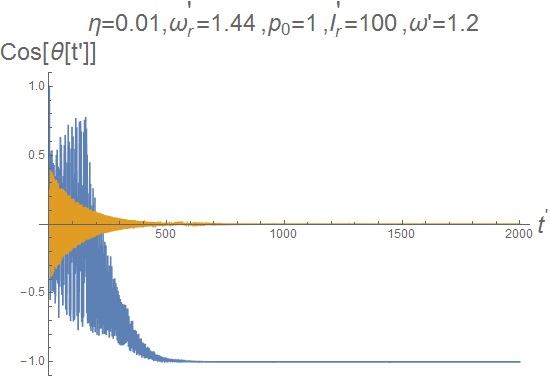}{a}
\includegraphics[width=80mm]{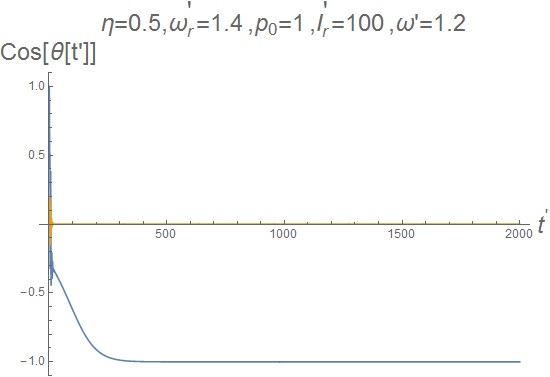}{b}
\caption{Switching dynamics of the magnetization in the coordinate frame of the magnet plotted together with its rotational dynamics for two sets of parameters. Blue line shows the projection of the normalized magnetic moment on the magnetic anisotropy axis. Yellow line shows spatial orientation of the torsional oscillator given by $\rho/(2\pi)$.} 
\label{dynamics1}
\end{figure}
\begin{figure}
\includegraphics[width=80mm]{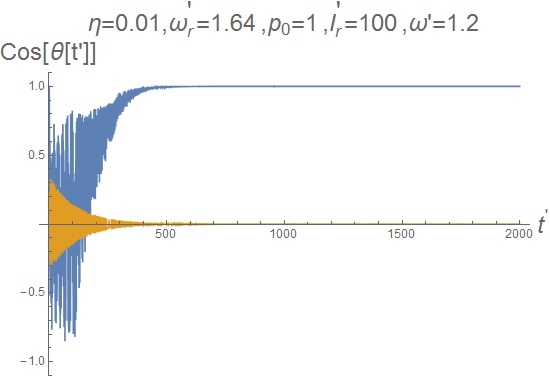}{a}
\includegraphics[width=80mm]{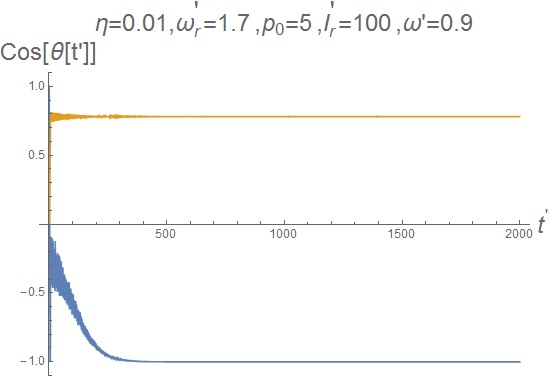}{b}
\caption{a) Dynamics of the magnetization and rotational dynamics of the oscillator showing no switching at long times for the chosen set of parameters. b) Switching of the magnetic moment accompanied by the permanent twist of the oscillator ($\rho \neq 0$) at the end of the field cycle for another set of parameters with $p_0 > p_c$.}
\label{dynamics2}
\end{figure}
In this Section we present results of the numerical study of the magnetization reversal governed by equations (\ref{theta-eq-rot-alpha}), (\ref{phi-eq-rot-alpha}), and (\ref{eq-rho-2}). In the numerical work we choose $\alpha = 0.01$ for the Landau-Lifshits-Gilbert damping parameter, which is in the right ballpark for single-domain magnetic particles \cite{Kalmykov-PRB2010}. As long as $\alpha$ is small, our results are not very sensitive to its exact value. They exhibit much greater sensitivity to the damping of the oscillations of the torsional resonator $\eta$. Strong mechanical damping provides deterministic ranges of parameters for which the switching occurs, while weak damping gives more stochastic dependence on parameters. For that reason we present numerical results for two cases: Weak mechanical damping, $\eta = 0.01$, and strong mechanical damping, $\eta = 0.5$. 

We apply a pulse of the rotating electric field, ${\bf E} = E_0(0,-\sin \omega' t', \cos \omega' t')$, of constant amplitude $E_0$ to the resonator shown in Fig. \ref{geometry}. The field is initially aligned with the electric polarization and then is rotated by $360^o$, which corresponds to a non-zero value of $\omega'$ within the time interval satisfying, $0 < \omega't' < 2\pi$ and zero $\omega'$ outside that time interval. This generates mechanical oscillations of the resonator and the ac magnetic field in the coordinate frame of the magnet. From practical point of view one only needs to switch the magnetization, preserving the orientation of the magnetization carrier in space. Static solutions of Eq.\ (\ref{eq-rho-2}) satisfy 
\begin{equation}\label{static}
\rho + p_0\sin\rho = 0
\end{equation}
The return of the oscillator to its initial, $\rho = 0$, orientation in space at the end of the field rotation is guaranteed if $\rho = 0$ is the only solution of Eq.\ (\ref{static}). It is easy to see that this occurs when the amplitude of the electric field satisfies $p_0 < p_c = 4.60334$. At high fields the oscillator stays twisted (that is, with $\rho \neq 0$) at the end of the field rotation if the field is not switched off or reduced below $p_0 = p_c$.

\begin{figure}
\includegraphics[width=75mm]{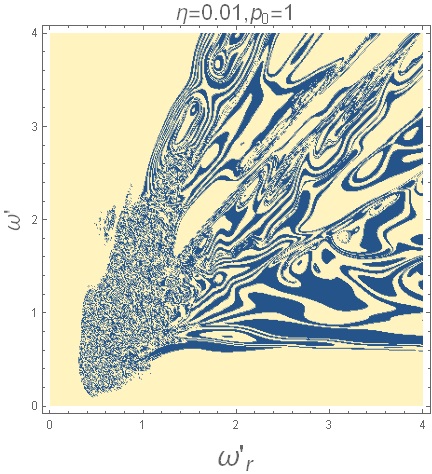}{a}
\includegraphics[width=75mm]{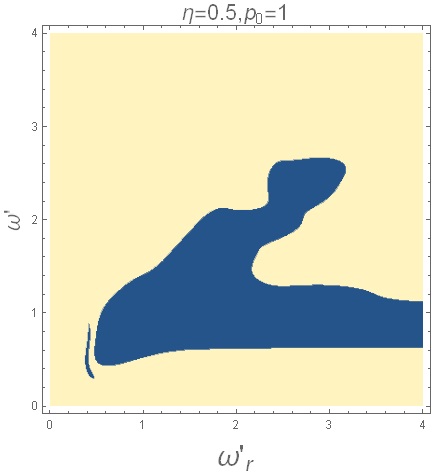}{b}
\caption{Switching phase diagram in the phase space $(\omega', \omega'_r)$. Areas of magnetization reversal are shown in blue (dark) color.}
\label{phasediagram-frequencies}
\end{figure}
\begin{figure}
\includegraphics[width=75mm]{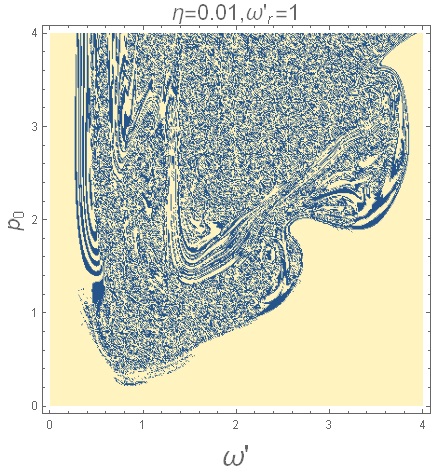}{a}
\includegraphics[width=75mm]{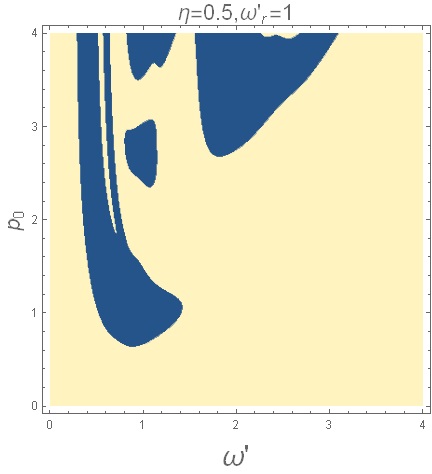}{b}
\caption{Switching phase diagram in the phase space $(p_0,\omega)$ representing the effect of the rotating electric field. Areas of magnetization reversal are shown in blue (dark) color.}
\label{phasediagram-field}
\end{figure}

Typical dynamics of the magnetization and the rotational dynamics of the oscillator are illustrated by Fig.\ \ref{dynamics1} and Fig.\ \ref{dynamics2}. The dynamics is sensitive to the ratio, $\omega'_r$, of the resonant frequency of the oscillator and the FMR frequency, and to the ratio, $\omega'$, of the frequency of the field rotation and the FMR frequency. It is also sensitive to the electromechanical factor $p_0$ that is proportional to the amplitude of the electric field $E_0$, and to the mechanical damping of the oscillator $\eta$. Switching vs non-switching is less sensetive to the parameter $\alpha$ when it is small, and has little sensitivity to the normalized moment of inertia of the oscillator $I'_r$ as long as it is large, meaning that under these conditions the damping term in Eq.\ (\ref{LL}) and the right-hand-side of Eq.\ (\ref{eq-rho-2}) contribute weakly to the dynamics of the system. In the numerical work we choose $I'_r = 100$. Magnetization reversal for strong and weak damping of the oscillator is illustrated in Fig.\ \ref{dynamics1}. For a different set of parameters chosen in Fig.\ \ref{dynamics2}a no reversal takes place at the end of the field-rotation cycle. When $p_0 > p_c$ the oscillator, in accordance with expectation, ends up in a twisted state with a non-zero asymptotic value of $\rho$, see Fig.\ \ref{dynamics2}b. 

Fig.\ \ref{phasediagram-frequencies} shows the switching phase diagram in the $(\omega', \omega'_r)$ plane. Weak mechanical damping, $\eta = 0.01$, results in a stochastic dependence on parameters, Fig.\ \ref{phasediagram-frequencies}a, which is a consequence of the high non-linearity of the dynamical equations. Strong damping, $\eta = 0.5$, gives more deterministic behavior characterized by a compact area of the magnetization reversal, Fig.\ \ref{phasediagram-frequencies}b. Both types of behavior exhibit thresholds on $\omega'$ and $\omega'_r$ required for the reversal. 

Switching phase diagram in terms of the parameters representing the rotating electric field, $(p_0,\omega')$, is shown in Fig.\ \ref{phasediagram-field}. Here again the weak mechanical damping of the oscillator results in a stochastic dependence of the switching on the parameters, which is illustrated by 
Fig.\ \ref{phasediagram-field}a, while strong damping gives more deterministic behavior, Fig.\ \ref{phasediagram-field}b. It shows islands in the $(p_0,\omega')$ plane that provide the required strength of the electric field and the speed of the field rotation. 

\section{Discussion} \label{discussion}
We have shown that a magnetic moment of a single-domain magnet that is allowed to perform torsional oscillations with respect to a fixed axis can be switched by the mechanical oscillations alone. This effect arises from the fact that in the reference frame of the magnet the rotation is equivalent to the magnetic field, thus making torsional oscillations equivalent to the ac magnetic field. It is well known \cite{CGC-PRB2013} that the ac magnetic field, even of small amplitude, is capable of  reversing the magnetic moment. The physics of the effect is related to the consecuitive absorption of photons that, having a non-zero angular momentum, change the projection of the spin angular momentum of the magnet. In that sense the mechanical rotations are equivalent to photons, with the spin angular momentum of the magnet absorbing the mechanical angular momentum associated with torsional oscillations.    

In this paper we propose to ignite the required mechanical oscillations by the time dependent electric field (voltage) applied to a multiferroic system or to a ferroelectric firmly coupled with a magnet. According to Fig.\ \ref{phasediagram-frequencies}, the switching of the magnetic moment occurs when the frequency characterizing the rate of change of the field, and the resonant frequency of the oscillator, are not very small compared to the frequency of the ferromagnetic resonance due to magnetic anisotropy (no external magnetic field is required). The latter is typically in the ballpark of a few GHz.  Nanomechanical oscillators of such frequencies are common nowadays. (Oscillation frequencies, $f = \omega/(2\pi)$, of hundreds of GHz have been reported in NEMS based upon carbon nanotubes \cite{Island}.) The effective magnetic field due to torsional oscillations of amplitude $\rho_0$ at a frequency $\omega = 2\pi f$  is of order  $h \sim \omega \rho_0/\gamma$. For $f$ of a few GHz and $\rho_0 \sim 0.1$ it is in the balpark of $10$mT, which is more than sufficient for magnetization reversal by the ac magnetic field \cite{CGC-PRB2013}, thus allowing for even weaker torsional oscillations than $\rho_0 \sim 0.1$ used in the estimate.  

The effect of the electric field on the oscillator is described by the electromechanical ratio $p_0 = PE_0/(I_r\omega_r^2)$. Fig.\ \ref{phasediagram-field} shows that the field pulse results in the magnetization reversal when $p_0$ is not too small compared to one. Estimating the electric polarization $P$ as $P_0V$ and the moment of inertia of the oscillator as $I_r \sim m r^2$, with $P_0, V, m$ and $r$ being the polarization density, the volume, the mass, and the size of the oscillator, respectively, one obtains that the voltage, $V_E \sim E_0 r$ required to operate the device must be of order $V_E \sim m\omega_r^2/P_0$. For $m \sim 10^{-20}kg$, $\omega \sim 10^{10}$s$^{-1}$, and $P_0 \sim 0.1 C/m^2$, this gives voltages in the ballpark of a few volt. Making such a device, while challenging, must be a doable task in the light of the recent progress in manufacturing NEMS. Notice that high quality factor of the nanooscillator that is required for their applications as sensors is not needed for the purpose of magnetization switching. In fact, as our study shows, strong mechanical damping of the oscillator produces more reliable results.

\section{Acknowledgements}
This research has been supported by the U.S. National Science Foundation through Grant No. DMR-1161571. Numerical work was supported in part by a grant of computer time from the City University of New York High Performance Computing Center under NSF Grants CNS-0855217, CNS-0958379 and ACI-1126113.

\end{document}